# Design of Reliable and Resilient Electric Power Systems for Wide-Body All-Electric Aircraft

Mona Ghassemi

ZEROES (Zero Emission, Realization of Optimized Energy Systems) Laboratory

Department of Electrical and Computer Engineering, The University of Texas at Dallas

Richardson, TX, USA

mona.ghassemi@utdallas.edu

**Abstract:** To achieve net-zero emissions by 2050, all-electric transportation could be a promising option. Among all the sectors that generate greenhouse gas emissions (GHGs) in the U.S., the transportation sector ranks first (29%) in contributing to those emissions. It is undeniable that electric vehicles are nearing maturity. Still, aviation is beginning to develop an electrified aircraft that can operate properly on commercial flights. More than 75% of aviation GHG emissions come from large aircraft, including narrow-body and wide-body aircraft; this situation will likely worsen considering a historic 4-5% annual air travel growth. Electrification of aircraft has led to the development of two types of aircraft: the more electric aircraft (MEA) and the all-electric aircraft (AEA). Aircraft electrification has given rise to two types of aircraft: more electric aircraft (MEA) and all-electric aircraft (AEA). A MEA replaces a subsystem, such as a hydraulically driven actuator, with an electric alternative. On the other hand, AEA comprises electrically driven subsystems and thrust power fully provided by electrochemical energy units (EEUs). This difference is substantial for wide-body AEA since the required thrust power is ~25 MW, and non-thrust demands add another 1 MW. This results in significant challenges for optimizing the AEA's electric power system (EPS) design, where the maximum component power density must be achieved by minimizing both mass and volume. The voltage level increment, in a few kV ranges and as a DC type (Medium Voltage Direct Current (MVDC)), is a feasible option for enhancing power transmission capability. As a result, the design of an MVDC electric power system (EPS) for wide-body AEA is needed. Since failures in EPS AEA may jeopardize the safety of passengers, reliability and resilience play vital roles. This chapter deals with designing reliable and resilient EPS for envisaged wide-body AEA. It presents an accurate load flow model for DC systems for aircraft EPS to determine the power flowing through aircraft cables for normal situation and all single contingencies. Then, optimal MVDC EPSs are discussed. To this end, a whole EPS for the envisaged wide-body AEA is introduced. The locations for the non-propulsion loads and EEUs are evaluated, the distances between loads and busbars are estimated, and flow studies are carried out. Various architectures are proposed and analyzed in terms of reliability, power density, power loss, and cost to obtain optimal architecture(s).

## 1. Introduction

Climate change and global warming issues are addressed by net-zero emissions goals in most countries. Net zero emissions, often referred to as "net zero," is the state where the amount of greenhouse gases emitted into the atmosphere is balanced by an equivalent amount removed, resulting in no net increase in atmospheric greenhouse gas levels. The top six greenhouse gas (GHG) emitters in 2019 were the United States, the United Kingdom, the European Union, China, Japan, and India, contributing around 55% of global GHG emissions [1]. Globally, in 2019, about 18% of GHG emissions came from the transportation sector, which is the second largest share of GHG emissions [1]. The discussed data are especially regarding 2019 (before the COVID-19 pandemic) since the pandemic halted the aviation industry more than other forms of transportation and transportation more than other economic sectors. In 2019, the aviation industry produced only 1.6% of GHG emissions globally, but its 4-5% growth will increase its share [1, 2]. Even by reaching a 2% efficiency improvement targeted by International Civil Aviation Organization (ICAO), an increase of 4% in demand growth will result in doubling emissions from aviation by 2050 compared to 2019 [3]. Therefore, the aviation industry must also decarbonize to realize the long-term net-zero emissions goal. To reach net zero, it's essential to both reduce greenhouse gas emissions and implement measures to remove existing gases from the atmosphere. Key strategies include 1) reducing emissions through i) transitioning to renewable energy by replacing fossil fuels with renewable energy sources like wind, solar, and hydroelectric power to generate electricity without emissions, ii) enhancing energy efficiency via improving the efficiency of buildings, transportation, and industrial processes to lower energy consumption and associated emissions, iii) electrification by shifting to electric vehicles and heating systems powered by renewable energy to reduce reliance on fossil fuels; 2) removing carbon from the atmosphere by i) natural solutions such as expanding forests, restoring wetlands, and adopting sustainable agriculture practices to enhance the natural absorption of $CO_2$, and 3) technological solutions by developing and deploying carbon capture and storage (CCS) technologies to capture $CO_2$ emissions from industrial sources or directly from the air and store them underground. Achieving net zero is crucial for mitigating the impacts of climate change. By balancing emissions with removals, we can stabilize global temperatures and reduce the frequency and severity of extreme weather events. The Paris Agreement emphasizes the need for countries to reach net zero emissions in the second half of this century to limit global warming to well below 2°C above pre-industrial levels. Transitioning to net zero requires significant changes across all sectors of the economy. Challenges include technological development, financial investment, policy implementation, and societal acceptance. It's also important to ensure that the transition is just and equitable, providing support for communities and industries affected by the shift to a low-carbon economy. In summary, net zero emissions represent a balance between the greenhouse gases we emit and those we remove from the atmosphere. Achieving this balance is essential for addressing climate change and requires comprehensive strategies encompassing emission reductions and carbon removal initiatives. Electric aircraft offer many advantages, including lower greenhouse gas emissions (assuming that all electricity is generated through renewable sources), lower energy consumption, more reliable electric subsystems that are easier to diagnose and prognosis and can be used only when necessary, and lower noise production than conventional aircraft [4-7]. As mentioned aircraft electrification involves integrating electric power into various aircraft systems, aiming to enhance efficiency, reduce emissions, and lower operational costs. This process ranges from incorporating electric components into traditionally powered aircraft to developing fully electric propulsion systems. Key Aspects of aircraft electrification are: 1) More Electric Aircraft (MEA): This approach focuses on replacing non-propulsive systems, such as hydraulic and pneumatic systems, with electric ones. The goal is to improve efficiency and reduce weight, leading to better fuel economy and lower emissions; 2) Hybrid-Electric Propulsion:

Combining traditional engines with electric propulsion systems, hybrid-electric aircraft aim to optimize energy use. For instance, gas turbines generate electrical power, which is then distributed to electric drives integrated into propulsion subsystems. This configuration can lead to better energy management and reduced fuel consumption; and 3) All Electric Aircraft (AEA): These aircraft rely entirely on electric power for propulsion, eliminating the need for conventional fuel. Advancements in battery technology and electric motors are crucial for the development of fully electric aircraft. Companies like Collins Aerospace are at the forefront of this innovation, envisioning a future with aircraft that consume less fuel, emit less carbon, and have reduced maintenance costs. Note that four main types of energy exist in aircraft: hydraulic, mechanical, electrical, and pneumatic. By replacing a subsystem such as a hydraulically driven actuator with an electrically driven alternative, an MEA can be achieved. But, in order to realize an AEA, besides all the subsystems, the thrust power must also be fully provided electrically, for example, via electrochemical energy units (EEUs). Hybrid gas-electric propulsions can be considered as MEA. Challenges in aircraft electrification are i) energy density: current battery technologies offer limited energy density compared to traditional aviation fuels, restricting the range and payload capacity of electric aircraft, ii) thermal management: efficiently managing the heat generated by electric systems is essential to ensure safety and performance, and iii) certification and regulatory framework: establishing standards and regulations for electric aircraft is an ongoing process, requiring collaboration between manufacturers and aviation authorities. In this regard, some recent developments are i) urban air mobility (UAM): companies are developing electric vertical takeoff and landing (eVTOL) aircraft to revolutionize urban transportation. For example, Vertical Aerospace's VX4 air taxi aims to reduce an hour-long car journey to just 11 minutes, cruising at 150 mph; ii) regional electric aircraft: startups like Elysian are designing battery-powered jets capable of carrying 90 passengers up to 500 miles on a single charge, rethinking airplane design to maximize current battery technology efficiency, and iii) hydrogen-electric propulsion: companies such as ZeroAvia are developing hydrogen-electric powertrains that enable aircraft to fly without emitting carbon, only producing water vapor. Their ZA600 powertrain is designed for 19-seat aircraft, with plans to scale up for larger planes. The path to widespread aircraft electrification involves overcoming technical challenges and establishing regulatory frameworks. However, ongoing research and development efforts are paving the way for a more sustainable and efficient future in aviation. In this regard, despite the favorable ratings of the MEA and AEA, state-of-the-art electric propulsion systems have low specific power and are not as viable as conventional aircraft [8]. Compared to traditional aircraft like Boeing 747, Boeing 737, and Boeing 707, MEA models, such as Airbus 380, Boeing 787, and NASA STARC-ABL, require much electric power. $P_{el}/P_{th}$ can be calculated and considered as the electrification rate where $P_{el}$ is the consumed electric power and $P_{th}$ is the maximum thrust power during take-off. The above ratio does not include the auxiliary power unit (APU). There are 550/868 seats on the Airbus 380, and its hydraulic components are partly electrified. In Airbus 380, 115 Vac is the highest voltage level, 600 kVA is its total electric power, and $P_{el}/P_{th}$<0.2%. Many pneumatic and hydraulic systems were replaced with electric systems in Boeing 787-8 (242/410 seats). For example, instead of substantial pneumatic loads like wing ice protection systems and environmental control systems (ECS), power electronics and electrical machines with compressors were replaced in Boeing 787-8. The total electrical power of Boeing 787-8 is 1 MVA; 235 Vac and ±270 V (540 Vdc) are its highest voltage levels; and $P_{el}/P_{th}$<1.5%. This ±270 Vdc is the highest power distribution voltage in existing aircraft. Siemens, Rolls-Royce, and Airbus collaborated on the E-FAN X (70 seats). Four engines power its propulsion. An electric motor of 2 MW powers one of the propulsion engines and is fed by an EPS of 3 kVdc ($P_{el}/P_{th} \approx 25\%$). NASA's Single-aisle Turboelectric AiRCraft with Aft Boundary Layer propulsion (STARC-ABL) (154 seats, single-aisle) has 1 kVdc EPS and is expected to have its first test flights in 2035. It has a parallel hybrid propulsion system driven by two motors fed by two generator-rectifier systems coupled to turbofans. The electric drive part has approximately 2.61 MW of power ($P_{el}/P_{th} \approx 30\%$). N3-X (297/330 seats, twin-aisle passenger aircraft) is another NASA project with a maiden flight possible in 2040 that will utilize superconductivity and DC networks to reduce weight [8, 9]. The thrust is provided by 14 distributed electric drive systems (14x1.56=24.96 MW). EPS voltage levels and power values for main commercial MEA and ongoing MEA/AEA projects were summarized above. It is important to remember that the maximum amount of power is usually required for takeoff and thrust in the aircraft. A great deal of potential exists in optimizing electrical systems since electrical distribution and protection comprise nearly 30% of the entire system's mass [11]. An electrified aircraft's electric power system (EPS) must deliver high power and be low in mass to achieve its advantages. Aircraft cables are essential components of aircraft EPS that transmit power from one node to another. To make MEA and AEA financially viable, it is essential to reduce the weight of the power transmission system while keeping it as efficient as possible. This formula can be used to determine the efficiency of a cable: $\eta = P_{Del}/(P_{Del} + P_{Loss})$, where $P_{Del}$ is the power delivered by the cable at its receiving end and $P_{Loss}$ stands for the ohmic loss. It is possible to increase a cable's current or voltage to enhance its power transmission capability. Increasing current requires a heavier conductor, which adversely affects system mass. In addition, protecting and controlling large operating currents requires larger interruption devices. As a result, the voltage level increment is a feasible option. However, the relationship between system mass and voltage level is not monotonic [12]. A higher voltage level, for example, requires thicker insulation. Adding more insulation increases system weight (but is acceptable over the current increase in weight). Raising the voltage level from 540 V to 4.8 kV can save 1.4 tons of cable conductor mass on a 2.6 MW single-aisle turboelectric aircraft design [13]. Based on NASA's study, a maximum mass saving occurs at a 6 kV voltage level, assuming the entire electrical system is operated under a 1 atm pressurized cryogen [14]. For the N3-X design, ±2 kV was the optimal voltage [14].

## 2. Electric Power System Architectures for AEA

The NASA N3-X [15] is a double-aisle, wide-body turboelectric aircraft, a type of MEA, that can carry 297/330 passengers. A turboelectric system, including two wingtip-mounted turboshaft engines that drive two cryogenically cooled superconducting electric generators, provides generation in turboelectric N3-X. At takeoff, a minimum of 25 MW is required, generated by fourteen motor/fan sets, each rated at 1.785 MW and distributed throughout the aircraft. Considering a single-engine failure scenario, each power plant (one engine/two generators) is sized to provide 25 MW. Compared to turboelectric N3-X aircraft, all-electric N3-X aircraft introduced in [16, 17] lack engines and use four electrochemical energy units (EEUs) to supply their EPS. Keeping distributed propulsion from the NASA design mentioned above, Fig. 1 illustrates the schematic of N3-X aircraft EPS designed to be all electric, where "M" stands for motor. "R" and "L" stand for motors placed on the right and left sides of the aircraft, respectively. Fig. 1 does not include APU, sensors, etc., and only provides a simplified schematic to illustrate the propulsion section that consumes the most power and to illustrate the non-propulsive demands such as ECS, WIPS, e-taxi system, actuation systems, hotel and gallery loads, lighting, and avionics. Studies carried out by NASA showed that a voltage level of 4.5-12kVdc is required to sustain thrust power during take-off [14]. In [16, 17], three ±5 kVdc (10 kVdc) power system architectures were introduced for the propulsion system. They were compared to considering the architectures as AC with a voltage of 10 kVac EPSs with the same architecture. Those three MVDC EPSs are discussed below. Similarly, a bipolar ±0.5 kVdc primary supply bus can be suggested for non-propulsive loads, and a 28 Vdc system can be considered for providing avionics like that already present in conventional aircraft, and commercialized MEAs such as Boeing 787.

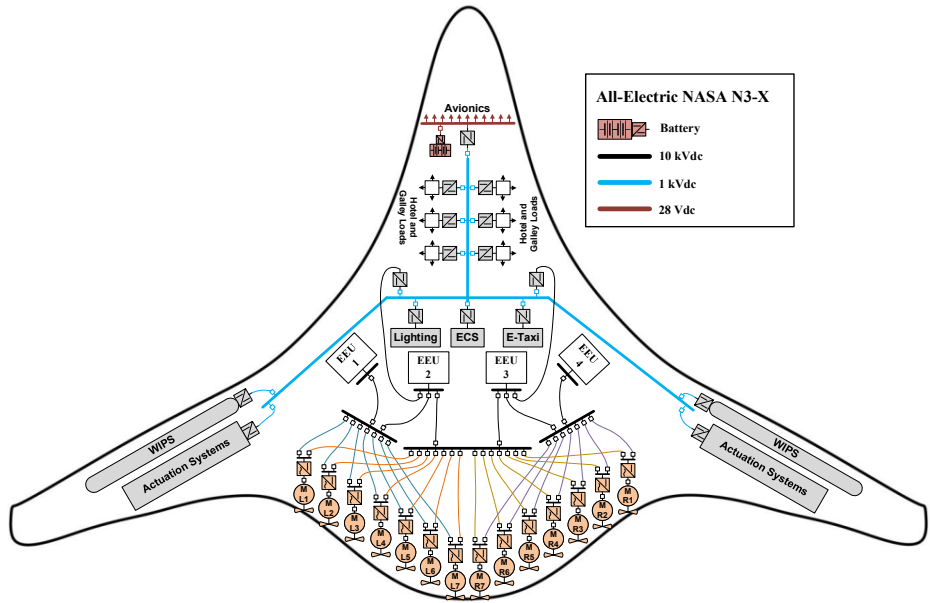

Fig. 1. Eelectric power system proposed for N3-X NASA aircraft, imagined as all-electric [16, 17].

The thrust power required for takeoff is assumed to be 25 MW. An MVDC EPS of ±5 kVdc is proposed to provide fourteen distributed motors, each of which is assumed to consume 1.785 MW of constant power during take-off. [16, 17] propose three propulsion architectures for the electric power system of NASA N3-X aircraft, considered to be all-electric. Fig. 2 shows Architecture #1, Fig. 3 shows Architecture #2, and Fig. 4 shows Architecture #3. Four EEUs, a combination of supercapacitors, batteries, and fuel cells, produce thrust power to supply fourteen motors in all architectures. The proposed architectures can sustain power flow to motors under normal situation and in all single contingency conditions even in the event of a severe failure such as losing an EEU. Although Figs. 2-4 depict DC architectures; all proposed architectures can be viewed as AC EPSs, where single-line representations may be shown for proposed AC propulsion sections. According to [16, 17], the EPSs in Figs. 2-4 satisfy power flow and bus voltage requirements under normal situation as well as under all single contingencies. All single contingencies in each architecture were studied in [16, 17], and the single contingencies causing the most severe conditions were reported in [16, 17]. We assumed single contingencies for busbars, cables (branches), and EEUs. The following shows that losing a busbar or EEU often leads to the most severe consequences. If, for example, we lose bus #5 in architecture #1, EEU1 is lost, the connections between the motors connected to bus #5 are lost, and cables 1-5 and 2-5 are lost. However, we still consider this a single contingency: losing one busbar. The designed

EPS maintains power flow to motors. In EEUs, the most severe condition was assumed to be losing a complete EEU in the case of an n-1 contingency. For example, when EEU2 is lost, the EPS no longer receives power from it. Nevertheless, in architecture #1, bus #2 is still available so that power can flow from bus #5 to bus #2, then from bus #2 to bus #6 to power motors.

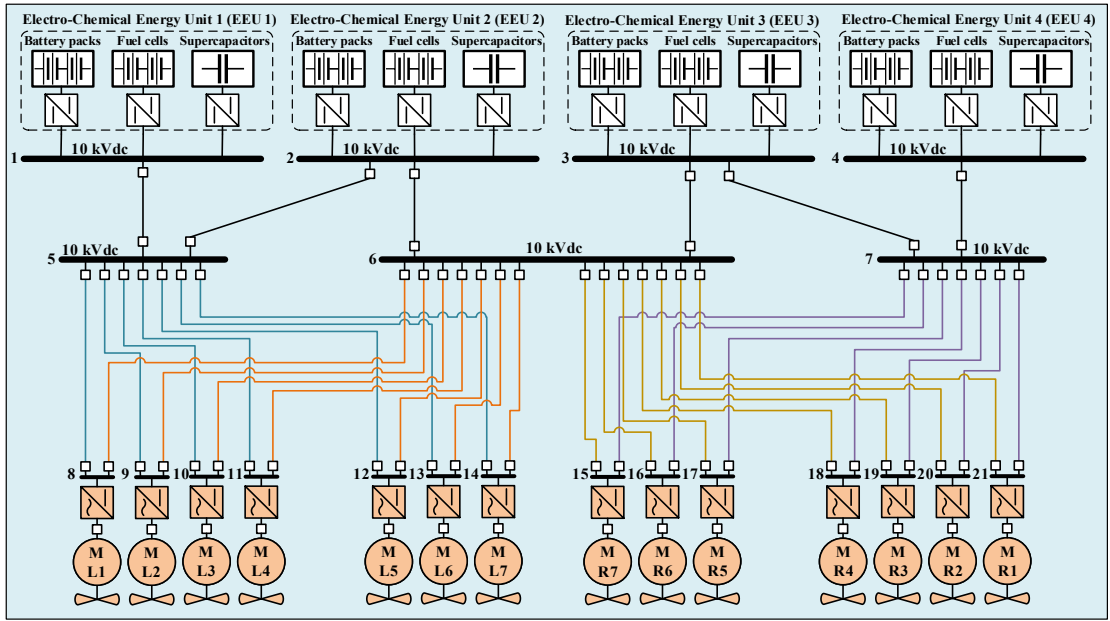

Fig. 2. Propulsion section of electric power system for N3-X NASA aircraft imagined as all-electric, Architecture #1 [16, 17]. MR1: motor #1, right wing, ML1: motor #1, left wing.

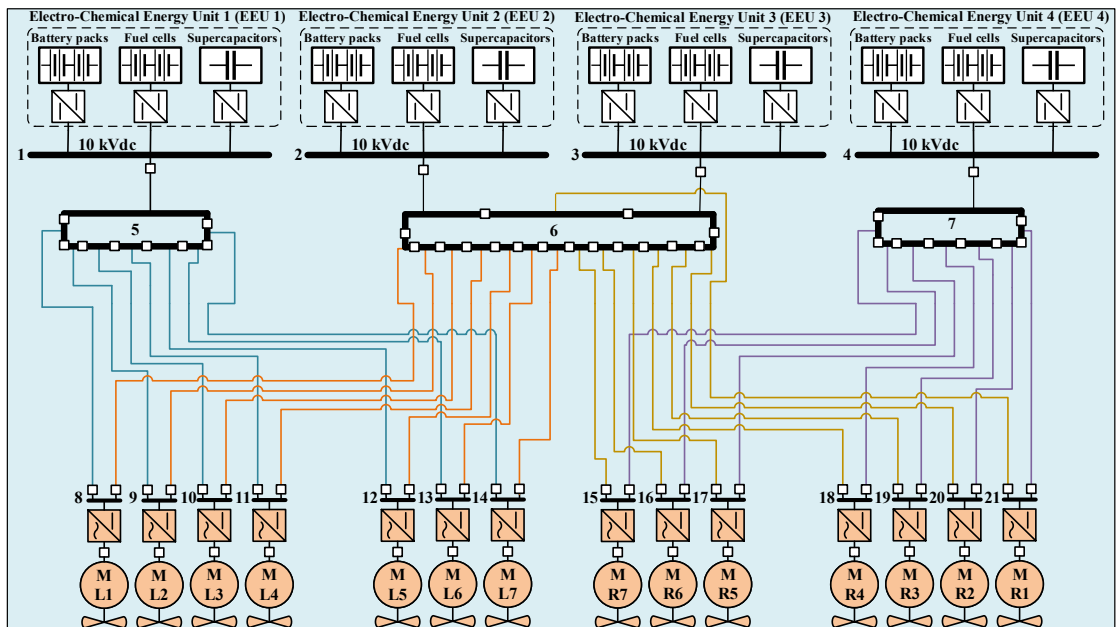

Fig. 3. Propulsion section of electric power system for N3-X NASA aircraft imagined as all-electric, Architecture #2 [16, 17].

The aircraft's EPS also supplies power to non-propulsive loads, which consume up to 5 MW. Non-propulsive loads, however, are almost turned off during take-off, so they were neglected. Circuit breakers (CBs) are needed in Figs. 2-4 to reconfigure and prevent a fault from propagating.

Depending on their construction, there are three types of CBs: mechanical, electrical, and hybrid (a combination of both). Suppose we ignore the CBs connecting the various types of EEUs to buses #1 to #4 and the CBs connecting buses #8-#21 in architectures #1 and #2 and #9-#22 in architecture #3 to the inverters/motors for architecture #1. In that case, 68 CBs are required, while for architectures #2 and #3, 64 and 72 CBs are needed, respectively.

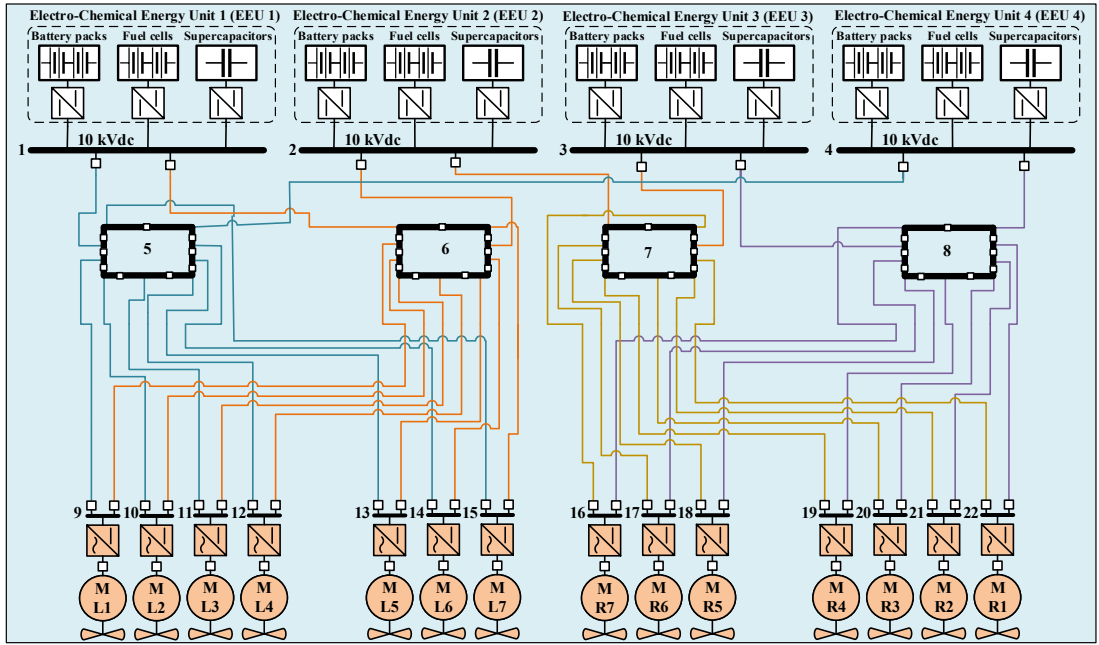

Fig. 4. Propulsion section of electric power system for N3-X NASA aircraft imagined as all-electric, Architecture #3 [16, 17].

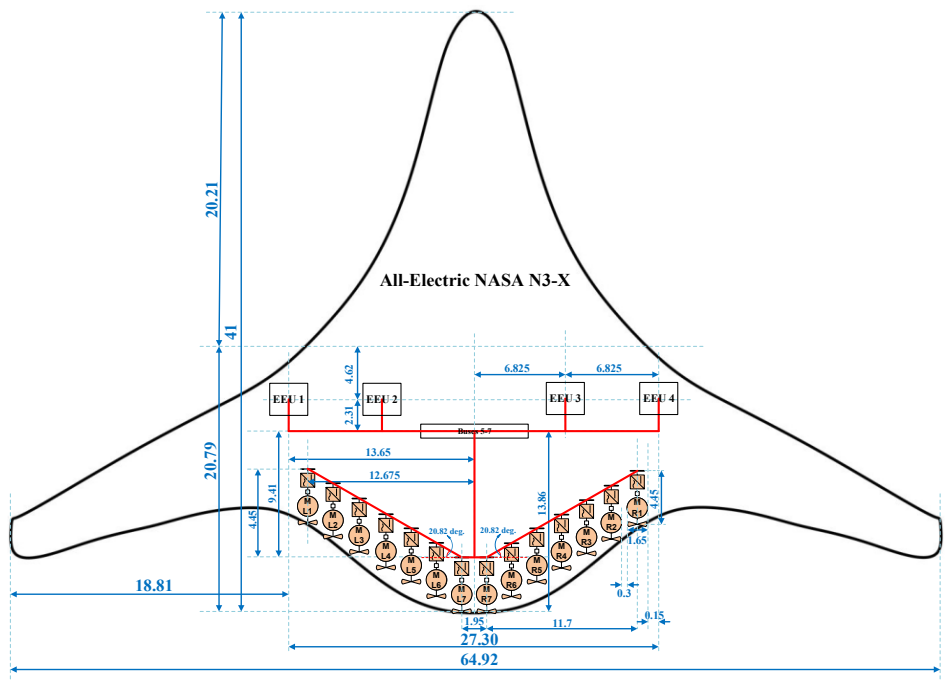

Fig. 5. Estimation of dimensions for envisaged all electric N3-X (all numbers are in meters).

A circuit breaker can fail without tripping, known as a circuit breaker failure (CBF); the three proposed architectures are all reliable with respect to all CBFs. For instance, it should be noted that in the event of a failure in bus 6 in architecture #1 and a CBF in one of the CBs in cable 2-6, the other CBs still act and EPS continues to work as usual. To understand the length of the aircraft cables in the architectures discussed above, Fig. 5, with red lines, suggests a distribution path for aircraft cables connecting motors, busbars, and EEUs. Based on the estimated dimensions in [16, 17] and shown in Fig. 5, the lengths of all cables in the proposed architectures were obtained. Using aircraft dimensions shown in Fig. 5 and cable lengths calculated for architecture #1, the total length of aircraft cables connecting buses 5-7 (busbars) to buses 1-4 (EEUs) is 68 meters. It is 50 m and 100 for architectures #2 and #3, respectively. All architectures also have 466 meters of aircraft cables that connect buses #9-#22 in architecture #3 and buses #8-#21 in architectures #1 and #2 to busbars. It should be noted that resilience, as defined for power grids [18, 19], does not mean for EPS architectures for wide-body AEA. In power grids, resilience is an umbrella term covering many factors. Resilience is the ability to prepare and cope with low probability, high-impact events in a robust manner so that their damaging impact is reduced effectively and/or the duration of the disruption is reduced, which includes the ability to absorb, adapt to, capture uncertainties, and recover quickly from such events. Low probability, high-impact events are often considered those resulting from natural hazards/disasters, viz., meteorological (e.g., severe storms, tropical cyclones, and extreme temperatures), hydrological (e.g., droughts, floods, and tsunamis), geological (e.g., volcanic eruptions and earthquakes), and biological (e.g., pandemics and epidemics) phenomena that inflict extensive damage and loss on the electric power sector. However, none of such natural disasters can be imagined for electric power systems in wide-body AEA. Instead, we expect to expand the reliability factor in single contingencies from considering only branches to the loss of a busbar or an EEU, which was done when designing architectures in Figs. 2-4.

## 3. Power Flow Solvers for DC Systems

It is necessary to develop power flow solvers for DC systems to analyze the three MVDC EPS architectures discussed in Section 2. However, there are many power flow solvers for AC systems, not many for DC systems. This section discusses two power flow methods for DC systems that were used to analyze the MVDC above EPS architectures in [16, 17, 20-22]: modified Z bus method and modified Monotone mapping. For the MVDC network, only constant power generation and loads are present. Power flow solvers are nonlinear and require an iterative method to solve.

***3.1 Modified Z-Bus Method:*** For a system with $n_t$ number of buses, $n_V$ is the number of voltage-controlled buses and $n_L$ is the number of load buses that are not voltage-controlled, then

$$n_L = n_t - n_V \tag{1}$$

The line current injected in each bus is as follows:

$$i_n = -\frac{p_n^{cst}}{v_n} - i_n^{cst} - y_n^{cst} v_n \tag{2}$$

where $v_n$ is the voltage of the bus $n$, and $p_n^{cst}$, $i_n^{cst}$, $y_n^{cst}$ are the constant-power, constant-current, and constant-admittance respectively. For a DC system, only constant power generators and loads are present in the system, so Eq. (2) becomes

$$i_n = -\frac{p_n^{cst}}{v_n} \tag{3}$$

By applying Kirchhoff's current law, we get

$$i_n = \sum_{k \in n_t} y_{nk}(v_n - v_k) \qquad (4)$$

From Eqs. (3) and (4),

$$-\frac{p_n^{cst}}{v_n} = \sum_{k \in n_t} y_{nk}(v_n - v_k) \qquad (5)$$

which can be written as

$$y'_n v_n = \sum_{k \in n_L} y_{nk} v_k + \sum_{k \in n_V} y_{nk} v_k - \frac{p_n^{cst}}{v_n} \qquad (6)$$

If we define the first term as $k_n$

$$k_n = \sum_{k \in n_L} y_{nk} v_k \qquad (7)$$

we can rewrite Eq. (6) as

$$v_n = \sum_{k \in n_L} \frac{y_{nk}}{y'_n} v_k + \frac{k_n}{y'_n} - \frac{p_n^{cst}}{y'_n v_n} \qquad (8)$$

For the implementation of the power flow problem, $n_L \times n_L$ matrix is formed from matrixes $y$ and $y'$ where its diagonal elements ($Y_{nn}$) are $y'_{n+nv}$ and off-diagonal elements ($Y_{nk}$) are $y_{n+nv,k+nv}$. We also need a matrix $v = [v_1 \ldots v_{nL}]^T$, $p_n^{cst} = [p_1 \ldots p_{nL}]^T$ and $k = [k_1 \ldots k_{nL}]$. Finally, a diagonal matrix diag($v$) is defined as follows

$$\text{diag}_i(v) = \begin{cases} v_n & i = n \\ 0 & i \neq n \end{cases} \qquad (9)$$

In an iterative process, the power flow problem can be solved as follows.

$$v^{t+1} = f(v^t) \qquad (10)$$

where $t$ means $t$-th iteration and the function $f(v)$ can be defined as

$$f(v) = Y^{-1}(k + diag^{-1}(v)p) \qquad (11)$$

For the checking existence, convergence and uniqueness of the solution matrix $r = Y^{-2}k$ is defined where $\underline{r} = \min_n |r_n|$ and $\overline{r} = \max_n |r|$. If we define a set $C_R = \|v - r\|_q \leq R$ for $R$ and $q \geq 1$ where according to [11], $R$ is the radius of $l_q$ norm ball $C_R$ and $\underline{R} \leq R \leq \overline{R}$.

$$R \in (\underline{R}, \overline{R}) = \left( \frac{\underline{r} - \sqrt{\underline{r}^2 - 4\alpha}}{2}, \quad \underline{r} - \sqrt{\alpha} \right) \qquad (12)$$

where $r = Y^{-2}k$, $\alpha = \|Y^{-1}\|_q \|p\|_q$, and $\underline{r}^2 \geq 4\alpha$.

**3.2 Modified Monotone Mapping Method:** From Eq. (6), we can write by substituting $u_n = v_n^2$

$$u_n = \sum_{k \in n_L} \frac{y_{nk}}{y'_n} \sqrt{u_n u_k} + \frac{k_n}{y'_n} \sqrt{u_n} - \frac{p_n^{cst}}{y'_n} \qquad (13)$$

Eq. (13) is solved by defining the $n_L$-length vector $u$ and the mapping $f$, where $u = f(u)$ and entries of $f$ are given as follows.

$$f_n(u) = \sum_{k \in n_L} \frac{y_{nk}}{y_n'} \sqrt{u_n u_k} + \frac{k_n}{y_n'} \sqrt{u_n} - \frac{p_n^{cst}}{y_n'} \tag{14}$$

Therefore, the power flow problem entails solving iterations of

$$u^{t+1} = f(u^t) \tag{15}$$

## 4. Power Flow Analysis in AEA EPS

It should be noted that the minimum power required during takeoff is 25 MW, as mentioned before, which is consumed through 14 distributed fan sets, each rated at 1.785 MW. However, only 30%, i.e., 7.5 MW power, is required for cruising altitude. However, cable design faces significant thermal challenges because of limited heat transfer at cruising height (12.2 km) due to low pressure, as shown in [23-33]. To address these issues, several cables have been developed in [34-39] that introduce the novel concept of Micro Multilayer Electrical Insulation (MMEI) systems. These cables are designed to prevent partial discharges (PDs) and arcs while offering higher current density. The experiments also revealed that they can maintain high breakdown voltage even at low pressures, making them suitable for high-voltage applications in electric aircraft. Therefore, it is necessary to answer the critical question of which of two situations is the worst (takeoff or cruising height). Although the aircraft cables' currents flow at a higher rate during takeoff, they do not face a heat transfer challenge due to the convective heat transfer limited at low pressure; however, the current is lower at cruising height. To answer the above question, we need first to know the maximum current magnitude flowing through different branches, and for that, we need to do a power flow analysis for the proposed MVDC EPS architectures for each situation (the cruising time and takeoff). In [16, 17], the power flow results for takeoff are presented, and in [22], the power flow results for cruising time are reported. In this section, some significant results from these two situations are discussed. The power flow analysis is carried out under normal situation and under all single contingencies for the proposed all electric N3-X NASA aircraft EPSs. Throughout all case studies, the power flow met the conditions outlined in Equation (12). The solution resulting from the solvers discussed in Section 3 is also unique. It is assumed that aircraft EPS is for steady-state during takeoff and that it uses four EEUs to provide the 25 MW thrust required at a voltage rating of ± 5 kVdc with 1.785 MW constant power consumption per motor. Buses 2 (EEU2) and 3 (EEU3) are assumed to be constant power generators, where EEU2 and EEU3 are sized to produce around 8.3 MW (25 MW/3). As long as all EEUs can supply EPS, EEU2 and EEU3 will work at 80% of their capacity (6.664 MW). In the event that an EEU fails, we assume EEU2 and EEU3 work at 100% capacity (around 8.3 MW). Buses 1 and 4 are both voltage controlled buses since in some circumstances, the aircraft's EPS may be split into two separate ones. Therefore, EEU1 and EEU4 provide the remaining power and the power loss. Additionally, if either EEU1 or EEU4 fails, the other would function as the only voltage-controlled bus. Based on these assumptions, all EEUs have the same capacity (~ 8.3 MW). Some power flow analysis results for architecture #1, shown in Fig. 2, are presented in Tables I and II for normal situation. From Table I, the maximum voltage drop is < 0.03%. Table II presents the current magnitude flowing branches under normal condition. Maximum loading percentages (MLP) for all aircraft cables/branches ($I_{max}$/ampacity) under normal situation are <56%. As a result, the architecture proposed in Fig. 2 satisfies both loading percentage requirements and voltage drop under normal situation. Under the worst single contingencies, when EEU2 fails, the maximum current is around 931 A flowing through aircraft cables 1-5.

TABLE I: Power Flow Results for DC under Normal Situation

| bus # | 1 | 2 | 3 | 4 | 5 | 6 | 7 |
|---|---|---|---|---|---|---|---|
| V (kV) | 5.0000 | 4.9997 | 4.9997 | 5.0000 | 4.9997 | 4.9995 | 4.9997 |
| Bus # | 8 | 9 | 10 | 11 | 12 | 13 | 14 |
| V (kV) | 4.9987 | 4.9987 | 4.9988 | 4.9989 | 4.9990 | 4.9991 | 4.9992 |
| bus # | 15 | 16 | 17 | 18 | 19 | 20 | 21 |
| V (kV) | 4.9992 | 4.9991 | 4.9990 | 4.9989 | 4.9988 | 4.9987 | 4.9987 |

TABLE II: Power Flow Results for DC under Normal Situation

| bus # from | to | Current (A) | bus # from | to | Current (A) |
|---|---|---|---|---|---|
| 1 | 5 | 583 | 6 | 12 | 79 |
| 2 | 5 | 110 | 6 | 13 | 77 |
| 2 | 6 | 556 | 6 | 14 | 75 |
| 3 | 6 | 556 | 6 | 15 | 75 |
| 3 | 7 | 110 | 6 | 16 | 77 |
| 4 | 7 | 583 | 6 | 17 | 79 |
| 5 | 8 | 96 | 6 | 18 | 80 |
| 5 | 9 | 97 | 6 | 19 | 81 |
| 5 | 10 | 97 | 6 | 20 | 82 |
| 5 | 11 | 98 | 6 | 21 | 82 |
| 5 | 12 | 100 | 7 | 15 | 104 |
| 5 | 13 | 101 | 7 | 16 | 101 |
| 5 | 14 | 104 | 7 | 17 | 100 |
| 6 | 8 | 82 | 7 | 18 | 98 |
| 6 | 9 | 81.84 | 7 | 19 | 97 |
| 6 | 10 | 81.03 | 7 | 20 | 97 |
| 6 | 11 | 80.02 | 7 | 21 | 96 |

Architectures #2 and # 3 also meet voltage drop and line loading under normal situation and all single contingencies. Details can be found in [16]. The results obtained with architectures #1 and #3 under 5 kVdc show satisfactory performance. However, architecture #1 is better than architecture #3, taking into account the number of circuit breakers, cost, weight, complexity, and size compared to architecture #3. Power flow analysis in [22] for architecture #1 for cruising time shows that the maximum current flowing aircraft cables is 363 A. [16] also analyzed the power flow during normal situation and all single contingencies for NASA N3-X designed to be all electric using AC EPSs; compared to DC case, a voltage of 10 kVac, 60 Hz was considered instead of bipolar 5 kVdc. Whenever all EEUs supply the EPS, the slack bus is assumed to be bus 1, while voltage-controlled (PV) buses are buses 2-4 whose voltages and real power are set to 10 kVac and 6.25 MW, respectively. If a failure results in the loss of a voltage-controlled EEU, real power of other voltage-controlled EEUs will be set to 8.33 MW. If a failure results in the loss of bus 1 (the slack bus), bus 2 is assumed to be the slack bus. Proposed architecture #1's power flow analysis results for normal situation are given in Tables III-VII for two sets of Mazama/Poppy and Helens/Pansy conductors. Details about these conductors can be found in [40].

TABLE III: Power Flow Results for AC under Normal Situation (Helens/Pansy)

| bus no. | 1 | 2 | 3 | 4 | 5 | 6 | 7 |
|---|---|---|---|---|---|---|---|
| V (kV) | 10.0000 | 10.0000 | 10.0000 | 10.0000 | 9.9995 | 9.9991 | 9.9995 |
| Angle (rad) | 0 | -0.00014 | -0.00014 | ~0 | -0.00013 | -0.00017 | -0.00013 |
| bus no. | 8 | 9 | 10 | 11 | 12 | 13 | 14 |
| V (kV) | 9.9972 | 9.9972 | 9.9980 | 9.9980 | 9.9978 | 9.9978 | 9.9983 |
| Angle (rad) | ~0 | ~0 | ~0 | ~0 | ~0 | ~0 | -0.00011 |
| bus no. | 15 | 16 | 17 | 18 | 19 | 20 | 21 |
| V (kV) | 9.9983 | 9.99760 | 9.9976 | 9.9974 | 9.9974 | 9.9981 | 9.9981 |
| Angle (rad) | -0.00011 | ~0 | ~0 | ~0 | ~0 | -0.00010 | -0.00010 |

In both pairs of conductors, the maximum mismatch is less than 0.001 W. For Mazama/Poppy and Helens/Pansy, the power loss is 5.34 kW and 6.60 kW, respectively. In Table VII, we summarize MLP for normal situation; MLP for the worst single contingencies and the single contingency that the worst case scenarios will occur. The maximum mismatch for Architecture #2 and Architecture #3 is less than 0.001 W as well. The highest power loss is for architecture #2 (5.55 kW for Mazama/Poppy and 6.88 kW for Helens/Pansy). Architecture #3, when using Mazama/Poppy, has the lowest power loss (4.63 kW). Table VII summarizes the PF results for all architectures. Compared to DC architectures, AC architectures have higher total power losses and more loading on branches, as expected. The DC option is, therefore, preferred for wide-body AEA.

TABLE IV: POWER FLOW RESULTS FOR AC UNDER NORMAL SITUATION (HELENS/PANSY)

| bus # from | bus # to | Current (A) | Angle (rad) | bus # From | bus # to | Current (A) | Angle (rad) |
|---|---|---|---|---|---|---|---|
| 1 | 5 | 633 | -2.99 | 6 | 12 | 101 | 2.42 |
| 2 | 5 | 411 | 1.81 | 6 | 13 | 101 | 2.42 |
| 2 | 6 | 707 | 2.41 | 6 | 14 | 100 | 2.32 |
| 3 | 6 | 705 | 2.41 | 6 | 15 | 100 | 2.32 |
| 3 | 7 | 411 | 1.82 | 6 | 16 | 101 | 2.44 |
| 4 | 7 | 632 | -2.99 | 6 | 17 | 101 | 2.44 |
| 5 | 8 | 110 | 2.70 | 6 | 18 | 101 | 2.46 |
| 5 | 9 | 110 | 2.70 | 6 | 19 | 101 | 2.46 |
| 5 | 10 | 113 | 2.76 | 6 | 20 | 100 | 2.37 |
| 5 | 11 | 113 | 2.76 | 6 | 21 | 100 | 2.37 |
| 5 | 12 | 112 | 2.74 | 7 | 15 | 117 | 2.81 |
| 5 | 13 | 112 | 2.74 | 7 | 16 | 111 | 2.72 |
| 5 | 14 | 117 | 2.82 | 7 | 17 | 111 | 2.72 |
| 6 | 8 | 102 | 2.47 | 7 | 18 | 110 | 2.71 |
| 6 | 9 | 102 | 2.47 | 7 | 19 | 110 | 2.71 |
| 6 | 10 | 101 | 2.39 | 7 | 20 | 114 | 2.78 |
| 6 | 11 | 101 | 2.39 | 7 | 21 | 114 | 2.78 |

TABLE V: POWER FLOW RESULTS FOR AC UNDER NORMAL SITUATION (MAZAMA/POPPY)

| bus no. | 1 | 2 | 3 | 4 | 5 | 6 | 7 |
|---|---|---|---|---|---|---|---|
| V (kV) | 10.0000 | 10.0000 | 10.0000 | 10.0000 | 9.9995 | 9.9992 | 9.9995 |
| Angle (rad) | 0 | -0.00013 | -0.00013 | ~0 | -0.00012 | -0.00016 | -0.00012 |
| bus no. | 8 | 9 | 10 | 11 | 12 | 13 | 14 |
| V (kV) | 9.9977 | 9.9977 | 9.9983 | 9.9983 | 9.9981 | 9.9981 | 9.9986 |
| Angle (rad) | ~0 | ~0 | -0.00010 | -0.00010 | ~0 | ~0 | -0.00011 |
| bus no. | 15 | 16 | 17 | 18 | 19 | 20 | 21 |
| V (kV) | 9.9986 | 9.9980 | 9.9980 | 9.9978 | 9.9978 | 9.9984 | 9.9984 |
| Angle (rad) | -0.00011 | ~0 | ~0 | ~0 | ~0 | -0.00011 | -0.00011 |

TABLE VI: POWER FLOW RESULTS FOR AC UNDER NORMAL SITUATION (MAZAMA/POPPY)

| bus # from | bus # to | Current (A) | Angle (rad) | bus # from | bus # to | Current (A) | Angle (rad) |
|---|---|---|---|---|---|---|---|
| 1 | 5 | 629 | -3.04 | 6 | 12 | 99 | 2.41 |
| 2 | 5 | 383 | 1.88 | 6 | 13 | 99 | 2.40 |
| 2 | 6 | 694 | 2.39 | 6 | 14 | 98 | 2.29 |
| 3 | 6 | 693 | 2.39 | 6 | 15 | 97 | 2.29 |
| 3 | 7 | 383 | 1.89 | 6 | 16 | 100 | 2.43 |
| 4 | 7 | 628 | -3.04 | 6 | 17 | 100 | 2.43 |
| 5 | 8 | 111 | 2.71 | 6 | 18 | 100 | 2.44 |
| 5 | 9 | 111 | 2.71 | 6 | 19 | 100 | 2.44 |
| 5 | 10 | 115 | 2.77 | 6 | 20 | 98 | 2.34 |
| 5 | 11 | 115 | 2.77 | 6 | 21 | 98 | 2.34 |
| 5 | 12 | 114 | 2.75 | 7 | 15 | 120 | 2.82 |
| 5 | 13 | 114 | 2.75 | 7 | 16 | 113 | 2.73 |
| 5 | 14 | 120 | 2.83 | 7 | 17 | 113 | 2.73 |
| 6 | 8 | 101 | 2.45 | 7 | 18 | 112 | 2.72 |
| 6 | 9 | 101 | 2.45 | 7 | 19 | 112 | 2.72 |
| 6 | 10 | 99 | 2.37 | 7 | 20 | 117 | 2.79 |
| 6 | 11 | 99 | 2.37 | 7 | 21 | 117 | 2.79 |

TABLE VII: POWER FLOW SUMMARY FOR ARCHITECTURES FOR AC AND DC CASES

| | Architecture | Cable Conductor | Normal Condition | | | The Worst Case of Single Contingencies | | |
|---|---|---|---|---|---|---|---|---|
| | | | Loss (kW) | MLP (%) ($I_{max}$/Ampacity) | in Cable | MLP (%) ($I_{max}$/Ampacity) | in Cable | Contingency |
| dc | #1 | Helens(AAC-TW) AAC Pansy | 4.67 | 55.30 48.29 | 1-5 & 4-7 5-14 & 7-15 | 88.30 83.07 | 1-5 / 4-7 busbars-8/21 | Failure of EEU2/EEU3. Loss of a busbar |
| | #1 | Mazama(AAC-TW) AAC Poppy | 3.75 | 49.64 41.73 | 1-5 & 4-7 5-14 & 7-15 | 79.23 71.43 | 1-5 / 4-7 busbars-8/21 | Failure of EEU2/EEU3. When a busbar is lost. |
| | #2 | Helens(AAC-TW) AAC Pansy | 4.81 | 63.17 45.60 | 2-6 & 3-6 6-14 & 6-15 | 78.98 100.03 | 1-5 / 4-7 6-14 / 6-15 | Failure of EEU2/EEU3. Failure of EEU1/EEU4. |
| | #2 | Mazama(AAC-TW) AAC Poppy | 3.86 | 56.71 39.21 | 2-6 & 3-6 6-14 & 6-15 | 70.91 88.50 | 1-5 / 4-7 6-14 / 6-15 | Failure of EEU2/EEU3. Failure of EEU1/EEU4. |
| | #3 | Helens(AAC-TW) AAC Pansy | 4.15 | 33.60 42.30 | 2-6 6-15 | 61.31 83.06 | 2-6 busbars-9 | Failure of EEU1 Loss of busbar #9 |

| | | | | | | | |
|---|---|---|---|---|---|---|---|
| | #3 | Mazama(AAC-TW) | 3.30 | 30.17 | 2-6 | 55.09 | 2-6 | Failure of EEU1 |
| | | AAC Poppy | | 36.40 | 6-15 | 71.43 | busbars-9 | Loss of busbar #9 |
| ac | #1 | Helens(AAC-TW) | 6.60 | 66.98 | 2-6 | 154.26 | 3-6 | Loss of cable 2-6 |
| | | AAC Pansy | | 54.26 | 7-15 | 128.13 | 5-14 | Loss of cable 2-6 |
| | #1 | Mazama(AAC-TW) | 5.34 | 59.11 | 2-6 | 121.53 | 3-6 | Loss of cable 2-6 |
| | | AAC Poppy | | 48.04 | 7-15 | 100.53 | 5-14 | Loss of cable 2-6 |
| | #2 | Helens(AAC-TW) | 6.88 | 78.65 | 2-6 & 3-6 | 141.65 | 3-6 | Loss of EEU2 |
| | | AAC Pansy | | 58.74 | 6-14 | 122.52 | 6-14 | Loss of EEU1 |
| | #2 | Mazama(AAC-TW) | 5.55 | 70.04 | 2-6 & 3-6 | 120.78 | 3-6 | Loss of EEU2 |
| | | AAC Poppy | | 49.88 | 6-14 | 105.36 | 6-14 | Loss of EEU1 |
| | #3 | Helens(AAC-TW) | 5.76 | 42.69 | 4-8 | 76.46 | 4-8 | Loss of cable 3-8 |
| | | AAC Pansy | | 52.40 | 8-16 | 97.76 | 6-9 | Loss of cable 5-9 |
| | #3 | Mazama(AAC-TW) | 4.63 | 37.96 | 4-8 | 66.22 | 4-8 | Loss of cable 3-8 |
| | | AAC Poppy | | 45.17 | 8-16 | 84.07 | 6-9 | Loss of cable 5-9 |

It should be noted that in terms of net-zero aviation, hydrogen aircraft are another promising solution. Hydrogen-fueled combustion engines, especially those integrated with low-$NO_X$ combustors, represent a viable solution for reducing emissions in aviation. The ENABLEH2 program in Europe has demonstrated advancements in hydrogen combustion technologies that significantly reduce nitrogen oxide emissions, another critical factor in achieving net-zero aviation. Furthermore, hydrogen boasts a high energy density of approximately 120 MJ/kg, compared to conventional fuels like gasoline, which has an energy density of 44.4 MJ/kg. This higher energy density makes hydrogen a strong contender for long-haul flights, offering both sustainability and efficiency advantages over traditional jet fuel. As governments and organizations worldwide continue to support hydrogen-based solutions, hydrogen will likely play a central role in aviation's transition to zero emissions by 2050. However, using hydrogen as fuel in aircraft faces significant challenges, particularly in storing liquid hydrogen. Due to its lower density compared to conventional fuels, hydrogen requires much larger storage tanks, which complicates aircraft design and integration. Historically, hydrogen research for aviation dates back to the 1930s, with the first aero-derivative hydrogen gas turbine by the German engineer Hugo Junkers, which demonstrated the feasibility of hydrogen propulsion in aircraft. This was followed by military projects in the 1950s. During this decade, significant military projects involving hydrogen propulsion emerged. Notably, the U.S. National Advisory Committee for Aeronautics (NACA) modified the Martin B-57 bomber to run on hydrogen during the "Project Bee" initiative. This marked one of the first instances of a manned aircraft transitioning between conventional jet fuel and hydrogen. Additionally, the U.S. Air Force's interest in high-speed reconnaissance led to the Lockheed CL-400 "Suntan" project, which aimed to develop a hydrogen-fueled aircraft for high-altitude missions. In the 1980s and 1990s, niche projects like the Soviet Union's Tu-155 made pioneering strides by conducting flights powered by liquid hydrogen. By the 2000s, hydrogen fuel cells became a focal point, driven by concerns over climate change, with initiatives like NASA's Helios program demonstrating the potential of hydrogen fuel cells in high-altitude, long-endurance unmanned aerial vehicles. While these early endeavors highlighted the technological viability of hydrogen propulsion, the high costs and technical challenges delayed its widespread adoption. However, as the availability of oil continues to decline in the current decade, it becomes imperative that relying on these depletable resources will soon be unsustainable and economically unattractive. This will necessitate the search for alternative fuel sources to power aircraft in the future. Although the timing of this transition is uncertain, synthetic aviation fuels (synjet), liquid methane ($LCH_4$), and liquid hydrogen ($LH_2$) are among the most promising candidates. Other options like ethanol, methanol, and ammonia are less viable due to their low energy density. Synjet, derived from sources like coal or tar sands, can be seamlessly integrated into existing aircraft systems, making it a practical replacement for Jet A fuel. Liquid methane offers a higher energy density than Jet A but is less dense and energetic than hydrogen. Hydrogen, despite its challenges of low density and cryogenic requirements, offers the cleanest combustion and the highest energy potential. As a result, recent years have seen a renewed interest in hydrogen aviation research as a viable solution for reducing carbon emissions in the aviation industry. Hydrogen acts as both a fuel and a coolant, protecting engine and airframe components from heat damage. For example, liquid hydrogen absorbs heat from engine components, preventing overheating. During combustion, the expanded hydrogen reduces thermal loads on the aircraft structure. Hydrogen can be produced using renewable energy sources through the electrolysis of water. Electrolysis splits water molecules into hydrogen and oxygen using electricity, and if the electricity comes from renewable sources like wind or solar power, the hydrogen production process is essentially emission-free. And when burned, hydrogen emits only water vapor ($H_2O$) and nitrogen oxides ($NO_X$). This would eliminate harmful carbon dioxide gas ($CO_2$) and sulfur dioxide gas ($SO_2$) emissions, which are significant contributors to climate change and air pollution in conventional jet fuel combustion. This potential positions hydrogen as a highly promising option for the future of aviation propulsion systems.

The future of AEA is being shaped by advancements in battery technology, propulsion systems, and infrastructure. Several companies and research institutions are working on next-generation electric planes to make air travel greener, quieter, and more efficient. Several companies are developing regional, commercial, and urban air mobility (UAM) electric aircraft to replace fossil fuel-powered planes. Examples of regional and commercial electric aircraft are: Eviation Alice (Capacity: 9 passengers, Range: ~400 km (250 miles), Battery: Lithium-ion (820 kWh), Status: Successfully tested, aiming for commercial operations by 2027); Heart Aerospace ES-30 (Capacity: 30 passengers, Range: ~200 km electric-only, ~800 km hybrid-electric, Status: Backed by Air Canada & United Airlines, expected by 2028); and Airbus ZEROe (Hydrogen-Electric Hybrid) (Capacity: 100-200 passengers, Range: ~2000-4000 km, Technology: Hydrogen fuel cells + electric propulsion, Target: Launch by 2035). Examples of eVTOLs & UAM are: Joby Aviation eVTOL (Capacity: 5 passengers, Range: ~240 km (150 miles), Speed: 320 km/h, Target: Air taxi service by 2025; Lilium Jet (Capacity: 7 passengers, Technology: Ducted electric fans for vertical takeoff & landing (VTOL), Target: Urban and regional transport); and Archer Midnight (Capacity: 4 passengers, Range: 160 km (100 miles), Goal: Air taxi deployment by 2025).

Key technologies driving electric aircraft are: 1) next-generation energy storage including i) solid-state batteries (2-3x energy density vs. lithium-ion, increased safety (no thermal runaway), expected by 2030), ii) lithium-sulfur batteries (higher energy density & lower cost, lighter than lithium-ion, still in R&D phase), and iii) hybrid hydrogen-battery systems (hydrogen fuel cells for longer flights, expected to power mid-range aircraft); 2) high-voltage electric power systems including i) future aircraft are moving toward higher DC systems, like ±5 kV discussed above, to reduce weight and improve efficiency, and ii) silicon carbide (SiC) & gallium nitride (GaN) power electronics improve performance; 3) advanced electric propulsion including i) axial flux motors: more power-dense and lighter than traditional radial motors, ii) distributed electric propulsion (DEP): multiple small motors for better control and efficiency.

In this regard, challenges and solutions are: 1) battery energy density limitations (current Li-ion batteries offer only ~250-300 Wh/kg, far below jet fuel (~12,000 Wh/kg)), Solution: solid-state & lithium-sulfur batteries could triple energy density; 2) charging infrastructure & grid demand (charging large aircraft requires megawatt-scale power), Solution: megawatt charging systems (MCS) for fast turnaround times at airports; 3) weight vs. range tradeoff (more batteries increase weight but improve range), Solution: hybrid hydrogen-electric systems could increase range without excessive weight; 4) regulatory & certification barriers (current aviation regulations are designed for fossil-fuel aircraft), Solution: FAA, EASA, and ICAO are developing new certification standards for electric planes.

Future outlooks (2030 & Beyond) are by 2030: Short-haul & regional flights (500-1000 km) will go fully electric, by 2035: Hydrogen-electric aircraft will enable mid-range flights (1000-3000 km), by 2040: Large electric aircraft (100+ passengers) will become viable, by 2050: the aviation industry aims for net-zero emissions with fully electric and hybrid-electric fleets.

Companies leading the charge are Airbus (developing hydrogen-electric and hybrid aircraft), Boeing (researching hybrid-electric propulsion for commercial planes), Rolls-Royce (developing advanced electric propulsion systems, ZeroAvia (leading in hydrogen-electric aviation), Eviation, Heart Aerospace, Lilium, Joby, Archer (electric regional aircraft and eVTOL pioneers).

In summary, the future of all-electric aircraft is bright but faces technological and regulatory hurdles. Battery advancements, hydrogen fuel cells, and high-voltage electric propulsion will be critical to achieving sustainable, zero-emission air travel.

Hybrid-electric aircraft (HEA) combine electric propulsion with traditional fuel-based engines, reducing fuel consumption, emissions, and operating costs. These aircraft serve as a transitional solution between conventional jet-fuel aircraft and fully electric or hydrogen-powered planes. A hybrid-electric aircraft uses both an internal combustion engine (ICE) and an electric propulsion system. There are two main types: 1) parallel hybrid system where i) both fuel engines and electric motors drive the propellers or fans, and ii) the aircraft can switch between electric-only, engine-only, or combined mode. An example is Embraer's Energia Hybrid-Electric (E9-HE). 2) series hybrid system where i) the fuel engine acts as a generator, charging the batteries and providing electricity to electric motors, ii) the engine does not directly power propulsion, only the electrical system. An example is Boeing's hybrid-electric research with NASA.

Advantages of hybrid-electric aircraft are i) fuel efficiency: reduces fuel consumption by 30-50% compared to conventional aircraft; ii) lower emissions: cuts $CO_2$ and $NO_x$ emissions, helping aviation meet climate goals; iii) quieter operation: electric motors reduce noise pollution, especially during takeoff and landing; iv) range extension: overcomes battery limitations by using fuel as a backup.

Leading hybrid-electric aircraft projects are 1) Airbus E-Fan X (Retired) (Configuration: series hybrid-electric aircraft (2MW electric motor), Goal: To test hybrid propulsion for larger commercial jets, Status: Canceled in 2020, but findings contribute to Airbus ZEROe hydrogen research); 2) Boeing + NASA Electrified Powertrain Flight Demonstrator (EPFD) (Configuration: parallel hybrid system for commercial aircraft, Technology: High-power electric motors assist traditional jet engines, Goal: Improve fuel efficiency and enable hybrid propulsion for future Boeing jets); 3) Embraer Energia Hybrid-Electric (E9-HE) (Capacity: 9 passengers, Technology: hybrid-electric propulsion for short-haul flights, Status: Expected entry into service in the 2030s); 4) Heart Aerospace ES-30 (United Airlines Backed) (Capacity: 30 passengers, Range: Electric-only: 200 km, Hybrid-electric: 800 km, Status: Expected commercial launch by 2028); 5) Ampaire Eco Caravan (Hybrid-Electric Cessna) (Conversion of a Cessna Grand Caravan into a hybrid-electric aircraft, Fuel savings: 50% less than conventional models, Status: Flight-tested, aiming for commercial use soon).

Key technologies in hybrid-electric aircraft are 1) batteries & energy storage including i) Lithium-Ion Batteries (Current Tech): used in most hybrid-electric prototypes, ii) solid-state batteries (Future Tech): higher energy density, expected post-2030; 2) electric motors & power electronics where i) High-efficiency electric motors assist or replace traditional jet turbines and ii) silicon carbide (SiC) power electronics improve efficiency; 3) turbogenerators for power generation where small gas turbine generators act as range extenders in hybrid aircraft.

Challenges and future outlook are 1) battery limitations (Current batteries are too heavy for large aircraft, Solution: next-gen batteries (solid-state, lithium-sulfur) & hybrid fuel cell systems; 2) Infrastructure & Certification where i) airlines need charging and hybrid support infrastructure, and ii) aviation authorities (FAA, EASA) are working on hybrid-electric regulations.

Expected hybrid-electric adoption timeline is 2025-2030: regional hybrid-electric aircraft enter service, 2030-2035: hybrid propulsion used in larger commercial aircraft, 2040+: Long-haul hybrid-electric aircraft become viable.

In summary, hybrid-electric aircraft combine the benefits of electric and fuel propulsion, making them a practical solution for reducing aviation emissions before full-electric or hydrogen aircraft become mainstream.

A wide-body aircraft is a large airliner with two aisles, typically used for long-haul flights (e.g., Boeing 787, Airbus A350). While hybrid-electric and regional all-electric aircraft are advancing, an all-electric wide-body aircraft remains a major challenge due to battery limitations.

Key challenges for an all-electric wide-body aircraft are 1) battery energy density (Jet fuel has an energy density of ~12,000 Wh/kg, while current lithium-ion batteries are 250-400 Wh/kg; a fully electric Boeing 787-like aircraft would require batteries 40-50 times heavier than current fuel tanks); 2) power & propulsion requirements (wide-body jets require tens of megawatts of power for sustained flight, no existing battery system can supply this power for long-haul travel); 3) aircraft weight & design constraints (batteries are much heavier than fuel, and unlike fuel, they don't burn off during flight, meaning the aircraft must carry full battery weight throughout the journey); 4) Charging infrastructure & turnaround time (charging a massive battery pack could take several hours, compared to refueling a jet in minutes; and airports would need high-power charging infrastructure for electric wide-body aircraft); 5) regulatory & safety challenges (aviation safety agencies (FAA, EASA) require strict certification for new propulsion technologies; and battery fires or failures at high altitudes present new risks for commercial aviation).

Future technologies enabling all-electric wide-body jets are 1) next-generation batteries including i) solid-state batteries (potential for 2-4x energy density), ii) Lithium-sulfur & lithium-air batteries (higher capacity, but still in research phase), iii) Metal-air batteries (theoretical potential to rival jet fuel); 2) superconducting electric propulsion (cryogenic superconductors could reduce electrical losses & power large motors efficiently).

Hybrid & hydrogen-electric systems discussed above can be as interim solutions where hybrid-electric wide-body aircraft could emerge before fully electric models and Hydrogen fuel cells offer a potential long-haul zero-emission alternative.

Possible timeline for all-electric wide-body aircraft is 2030-2040: Small regional electric aircraft become commercially viable, 2040-2050: battery advancements may enable short-to-medium-haul electric jets (~100-150 passengers), 2050+: a fully electric wide-body aircraft *might* become feasible if battery technology advances significantly.

A 100% electric long-haul wide-body aircraft is still decades away, but advances in batteries, superconducting motors, and alternative fuels could make it a reality post-2050.

As discussed in this chapter book, MVDC power cables are a crucial component in the electrification of future wide-body aircraft, enabling efficient power transmission with reduced weight and losses. Here are some key aspects related to their development and application:

1. Why MVDC for all-electric aircraft?
   - Higher Efficiency: DC transmission reduces losses compared to AC due to the absence of reactive power and lower resistive losses at high voltages.
   - Weight Reduction: Eliminating bulky AC transformers and associated components can significantly reduce aircraft weight.
   - Power Density: MVDC systems allow for higher power transmission using thinner cables, which is critical for electric propulsion.
   - Simpler Power Conversion: Electric aircraft require multiple voltage levels, and MVDC simplifies conversion and distribution.

2. Key challenges of MVDC cables in aircraft are:
   - Dielectric Insulation: High voltages in a pressurized, high-altitude environment require advanced insulation materials to prevent breakdown.
   - Thermal Management: MVDC cables carry high currents, generating heat that must be efficiently dissipated to prevent thermal runaway.
   - Weight Constraints: Conductors and insulation materials must be lightweight while maintaining electrical and mechanical integrity.
   - Fault Protection: Unlike AC, MVDC lacks natural zero-crossing for current interruption, making fault protection more complex.
   - Flexibility and Durability: Aircraft cables need to withstand vibrations, mechanical stress, and extreme environmental conditions.

3. Potential Materials and Technologies
   - High-Temperature Superconductors (HTS): Can dramatically reduce resistive losses but require cryogenic cooling.
   - Advanced Insulation Materials: Polyimide, PEEK (polyether ether ketone), or nanocomposites for improved dielectric strength.
   - Carbon Nanotube (CNT) Conductors: Lighter and more conductive than copper, promising weight savings.
   - Shielding and EMI Suppression: Effective shielding materials to mitigate electromagnetic interference (EMI).

4. Applications in Future Wide-Body Aircraft
   - Electric Propulsion: Feeding high-power electric motors for propulsion.
   - More-Electric Systems: Replacing traditional hydraulic and pneumatic systems with electric actuators and environmental control systems.
   - Energy Storage and Distribution: Connecting high-energy-density batteries, fuel cells, or hybrid power sources.

5. Industry Efforts and Research
   - Boeing, Airbus, NASA, and research institutions are investigating MVDC architectures.

- Development of 1-10 kV MVDC aircraft power systems is underway for hybrid-electric and all-electric propulsion concepts.
- Integration of advanced circuit breakers, power electronics, and cooling strategies to support high-power MVDC distribution.

## 5. Conclusion

This chapter book discusses electric power systems for envisaged wide-body all-electric aircraft. The chapter started with trends towards net zero emissions and all-electric transportation as one of the highly potential solutions. Then, the focus went to all-electric aircraft. The estimated electric power required, especially for takeoff, was presented, and three MVDC power system architectures recently introduced in the literature were analyzed in detail. Two power flow solvers for DC systems were explained, and then, using those solvers, some power flow results for the architectures were discussed. This chapter book provides a framework for almost all the details one needs to consider when studying, developing, and dealing with different architectures for future wide-body AEA. A significant advantage of this chapter book is that it provides an overview of almost all the details that must be considered when studying, developing, and dealing with different kinds of electric power system architectures for wide-body AEA.